# Ultralow phase noise microwave generation with an Er:fiber-based optical frequency divider

F. Quinlan,* T. M. Fortier, M. S. Kirchner, J. A. Taylor, M. J. Thorpe, N. Lemke, A. D. Ludlow, Y. Jiang, C. W. Oates, and S. A. Diddams[2]

*National Institute of Standards and Technology, 325 Broadway, Boulder, CO 80305*
*Corresponding author: fquinlan@boulder.nist.gov;*

[2]*e-mail: sdiddams@boulder.nist.gov*

We present an optical frequency divider based on a 200 MHz repetition rate Er:fiber mode-locked laser that, when locked to a stable optical frequency reference, generates microwave signals with absolute phase noise that is equal to or better than cryogenic microwave oscillators. At 1 Hz offset from a 10 GHz carrier, the phase noise is below -100 dBc/Hz, limited by the optical reference. For offset frequencies > 10 kHz, the phase noise is shot noise limited at -145 dBc/Hz. An analysis of the contribution of the residual noise from the Er:fiber optical frequency divider is also presented.

Generating and distributing microwave signals with low phase noise is compelling for scientific applications such as remote synchronization at large facilities [1], local oscillators for fountain clocks [2], and very long baseline interferometry [3]. Fabry-Perot optical cavities can have quality factors approaching $10^{11}$, and can serve as ultrastable optical frequency references when a CW laser is locked to a cavity resonance [4]. Combining this optical frequency reference with the high fidelity frequency division of an optical frequency comb allows for the realization of microwave signals with phase noise surpassing state-of-the-art microwave oscillators [5, 6]. This was recently demonstrated with a Ti:sapphire-based optical frequency divider (OFD) locked to an ultrastable optical frequency reference, where phase noise of -104 dBc/Hz at 1 Hz offset from a 10 GHz carrier was reported [7]. Achieving this same level of performance in Er:fiber based OFDs would also be of interest, since the 1550 nm center wavelength of Er:fiber is advantageous for large-scale pulse distribution, and the lower cost and power requirements make fiber lasers more amenable to a compact, mobile microwave source. Recent measurements of Er:fiber-based OFDs have shown the potential of these lasers to generate microwaves with extremely low phase noise close-to-carrier. Residual noise of -120 dBc/Hz at 1 Hz offset from a 11.55 GHz carrier was shown, and noise at larger offset frequencies was limited to -130 dBc/Hz [8]. However, phase noise comparisons between an independent Er:fiber-based and Ti:sapphire-based system have been significantly higher, limited to -90 dBc/Hz at 1 Hz offset, and -120 dBc/Hz at larger offset frequencies on a 9.2 GHz carrier [5]. Here we report on an Er:fiber-based OFD capable of producing 10 GHz microwaves with absolute phase noise below -100 dBc/Hz at 1 Hz offset, limited by the optical frequency reference. For offset frequencies > 10 kHz, the phase noise is shot noise-limited at -145 dBc/Hz. As discussed in detail below, key to our demonstration of low phase noise is a 200 MHz repetition rate laser with a high-speed intracavity modulator, and low intrinsic relative intensity noise (RIN). The demonstrated phase noise meets or exceeds the 10 GHz phase noise from cryogenic microwave oscillators [9], and is more than 40 dB lower than 10 GHz room temperature oscillators at 1 Hz offset frequency [10].

The method by which an optical frequency comb may be used as an OFD is as follows. A mode of the comb is phase-locked to the optical frequency reference. Pulse formation via passive mode-locking enforces a constant relative phase among the laser modes. As a result, stabilizing one mode will transfer the stability of the reference to every optical mode of the comb [11]. The pulse repetition rate $f_{rep}$ may then be expressed as $f_{rep} = (v_{opt} - f_0 - f_b)/n$, where $v_{opt}$ is the frequency of the optical reference, $f_0$ is the comb offset frequency, $f_b$ is the difference frequency between the optical reference and a comb mode, and $n$ is an integer on the order of $10^5$-$10^6$. Thus $f_{rep}$ represents the frequency-divided optical reference. Accessing $f_{rep}$ is achieved by photodetecting the optical pulse train to generate a corresponding electrical pulse train. In the frequency domain, the train of electrical pulses is a series of discrete tones at harmonics of $f_{rep}$ from which any harmonic within the photodetector (PD) bandwidth may be selected as a microwave source.

The advantage of this technique is derived from the fact that the large frequency division from optical to microwave is accompanied by a large reduction in the phase noise power spectrum. The relation between the phase noise power spectra of the optical and microwave signals is given by $L(f)_{microwave} = L(f)_{optical}/N^2$, where $L$ is the single-sideband phase noise, and $N$ is the optical-to-microwave frequency ratio. For an optical frequency reference at 282 THz, the phase noise is reduced by ~90 dB when converted to a 10 GHz microwave signal. Assuming perfect fidelity division, optical phase noise below -10 $f^{-3}$ dBc/Hz of a state-of-the-art 282 THz optical frequency reference leads to phase noise below -100 $f^{-3}$ dBc/Hz on the derived 10 GHz carrier.

A schematic of the Er:fiber mode-locked laser, $f_0$ detection, and $f_b$ generation is shown in Fig. 1. A 40 cm length of highly-doped Er gain fiber is pumped by two polarization-multiplexed 980 nm diodes. A short free-space section includes waveplates and a polarization beam splitter (PBS) to excite nonlinear polarization rotation mode-locking. The free-space section also includes a 4 cm long LiNbO$_3$ phase modulator for cavity length stabilization [12]. The control bandwidth of the cavity length with the LiNbO$_3$ crystal is ~300 kHz, sufficient to suppress most seismic and acoustic disturbances to the mode-locked laser cavity. Considering the length of the Er-doped fiber, 44 cm of singlemode fiber (SMF), and the LiNbO$_3$ crystal, the estimated cavity dispersion is $2 \cdot 10^{-3}$ ps$^2$ at 1550 nm. The laser output is taken at the rejection port of the PBS, resulting in ~70 mW average power for pump power ~1W.

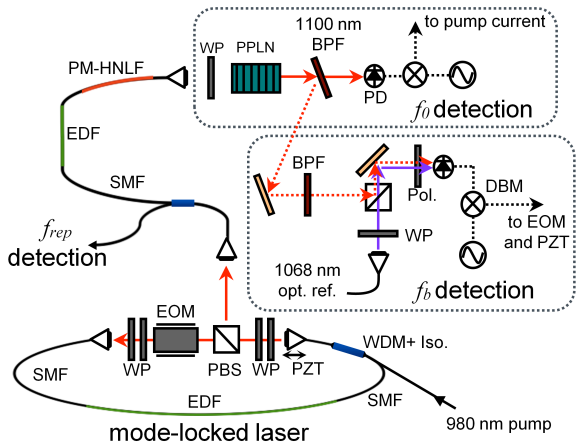

Fig. 1. (Color online) Er:fiber-based optical frequency divider. EDF, erbium-doped fiber; DBM, double balanced mixer; EOM, electro-optic phase modulator; WDM+Iso., wavelength division multiplexer/isolator hybrid; PZT, piezoelectric transducer; WP, waveplate; PPLN, periodically-poled LiNbO$_3$. Other symbols are defined in the text.

The laser output is split, with half the output used for $f_{rep}$ detection and monitoring, while the rest is first amplified to ~300 mW with an Er-doped fiber amplifier, then sent through 70 cm of polarization-maintaining, highly nonlinear fiber (PM-HNLF). After the PM-HNLF, the optical spectrum spans more than one octave, from 1050 nm to 2250 nm, suitable for $f_0$ detection in an f-2f interferometer. Frequency doubling of the 2200 nm light is accomplished in periodically-poled LiNbO$_3$. A 10 nm wide bandpass filter (BPF) centered at 1100 nm rejects light that does not contribute to the $f_0$ beat. A short piece of SMF after the PM-HNLF ensures overlap between the pulses at 1100 nm and 2200 nm for high signal-to-noise on $f_0$, typically 45 dB at a measurement resolution bandwidth of 300 kHz. The error signal generated from the $f_0$ beat is applied to the pump current.

The wavelength of the optical frequency reference used in this work is 1068 nm (282 THz). It is sent from a separate laboratory in the building to the Er:fiber comb via 300 m of noise-canceled optical fiber [13]. While the wavelength of the optical frequency reference is not directly accessible with the Er:fiber comb, it can be combined with the comb light rejected from the BPF in the f-2f interferometer to generate a beat of sufficient signal-to-noise, >30 dB at 300 kHz resolution bandwidth. High bandwidth locking of this beat signal is accomplished via feedback to the intracavity EOM, whereas low bandwidth/large dynamic range locking of $f_b$ is achieved via a piezoelectric transducer on one intracavity fiber launcher translation stage.

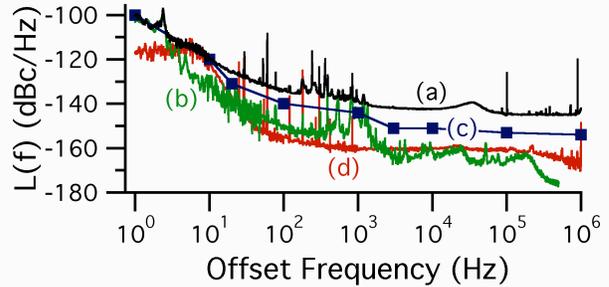

Fig. 2. (Color online) Single-sideband phase noise on a 10 GHz carrier. (a) Phase noise comparison between Er:fiber-based OFD and Ti:Sapphire-based OFD. (b) Optical phase noise comparison between the 1068 nm and 578 nm optical references, scaled to a 10 GHz carrier, assuming perfect fidelity division. (c) Phase noise contribution from the Ti:sapphire-based OFD. (d) Phase noise contribution of the measurement system's 10 MHz reference.

As the phase noise level of the generated microwaves is significantly lower than any commercially available source, the phase noise was measured by comparing against an independent Ti:sapphire OFD referenced to a stabilized 578 nm (518 THz) laser [4]. The repetition rates of the two systems are slightly offset to create a frequency difference of the 10 GHz harmonic of a few MHz. The photodetected pulse trains are filtered to select a harmonic near 10 GHz, amplified with low phase noise amplifiers, then combined in a double-balanced mixer. The beat signal at the output of the mixer is then compared to a quartz-based 10 MHz reference using digital phase comparison. The resulting phase noise comparison of the 10 GHz signals is shown in Fig. 2(a). At 1 Hz offset, the 10 GHz phase noise is below -100 dBc/Hz, limited by the phase noise of the optical references, shown in Fig. 2(b). Assuming both optical frequency references contribute equally, the phase noise of the Er:fiber-based OFD is ~-103 dBc/Hz at 1 Hz offset. The limit imposed by the optical references was determined by comparing the optical phase noise between the 1068 nm and the 578 nm signals, using the Ti:sapphire optical frequency comb to span the 236 THz frequency gap [7]. The optical phase noise level was then scaled to a 10 GHz carrier, assuming perfect fidelity division, by subtracting 90 dB from the optical phase noise. A separate microwave phase noise measurement between this Ti:sapphire OFD locked to the 578 nm reference and another Ti:sapphire OFD locked to the 1068 nm reference revealed a phase noise level given in Fig. 2(c). More details on the phase noise from the Ti:sapphire systems may be found in [7]. Thus the contribution of the Ti:sapphire OFD to this measurement system is only due to the 578 nm frequency reference, and only close to carrier. Phase noise of the 10 MHz reference is also shown in Fig. 2(d), and is seen to limit the phase noise measurement for offset frequencies 4 Hz - 10 Hz.

For offset frequencies greater than 10 Hz, the measured 10 GHz phase noise level may be due to a number of sources, including residual noise of the Er:fiber OFD; photodetector flicker noise [14], shot noise, and amplitude-to-phase noise conversion of the laser RIN; and fiber noise from the OFD to the PD. Here we highlight the contributions from noise sources directly related to the use of the Er:fiber-based OFD, namely shot noise, the residual noise of the Er:fiber OFD, and RIN. These noise contributions are shown in Fig. 3, along with the 10 GHz phase noise of Fig. 2(a). Curves (b) and (c) show the residual phase noise from $f_0$ and $f_b$, respectively, of the Er:fiber OFD. As with the optical comparison, 90 dB was subtracted from the measured data to scale to a 10 GHz carrier. Although the $f_0$ residual noise data was measured with slightly different lock conditions, it is nevertheless clear that this noise dominates the phase noise spectrum from ~1 kHz to 50 kHz. The residual noise of $f_b$ is significantly lower and does not contribute to the 10 GHz phase noise. A possible route to reduce the impact of the $f_0$ noise is to use the transfer oscillator technique of [15], effectively "mixing out" $f_0$. However, in our experience independently locking $f_0$ can reduce the laser RIN.

The calculated shot noise level is shown as the horizontal dashed line in Fig. 3. This level was calculated using the time invariant shot noise formula, with 4.36 mA of average photocurrent and power of the 10 GHz harmonic of -19.4 dBm. The power at 10 GHz was limited by saturation in the PD, which in turn bounded the achievable shot noise-limited signal-to-noise to -145 dBc/Hz. The shot noise level relates to the use of the Er:fiber-based system in that the PD saturation level depends strongly on the repetition rate of the pulse train on the PD, with higher repetition rates leading to higher output saturation power [16]. To date, $f_{rep}$ from Er:fiber-based optical frequency combs is limited to a few hundred MHz, whereas $f_{rep}$ of Ti:sapphire optical frequency combs has scaled as high as 10 GHz [17].

RIN can impact the 10 GHz phase noise through amplitude-to-phase conversion in photodetection [18]. The RIN of the Er:fiber laser was measured to be -125 dB/Hz at 1 Hz, decreasing to -145 dB/Hz at 1 kHz. Taking into account separate measurements on the amplitude-to-phase conversion coefficient on a similar PD, as well as shape of the RIN power spectrum, an upper limit can be placed on the projection of the RIN onto the phase noise. This "worst case" projection is shown Fig. 3(d). For the Er:fiber laser system shown here, the RIN minimally contributes to the 10 GHz phase noise. Possible sources of the phase noise level for offset frequencies 10 Hz to 1 kHz include PD flicker, and noise originating in the fiber link between the OFD and PD. The contribution of these sources is currently under investigation.

In conclusion, a Er:fiber-based optical frequency divider has been used to generate a 10 GHz signal with ultralow absolute phase noise. At 1068 nm, the optical reference wavelength was well outside the mode-locked spectrum of the Er:fiber laser. Locking to the optical reference was achieved by using part of the octave required for $f_0$ detection. Using this technique, it should be possible to use virtually any optical frequency reference from 1 μm to 2 μm in conjunction with an Er:fiber OFD for low noise microwave generation. It is worth emphasizing that although the phase noise has been characterized on a 10 GHz carrier, *any* harmonic of the 200 MHz repetition rate could be used as a ultralow phase noise microwave source.

We thank L. Nugent-Glandorf and W. Swann for their comments on the manuscript, M. Diddams for construction of the erbium fiber amplifier, and M. Hirano of Sumitomo Electric Industries for use of the PM-HNLF. Financial support is provided by NIST. F. Quinlan is supported as an NRC/NAS postdoctoral fellow.

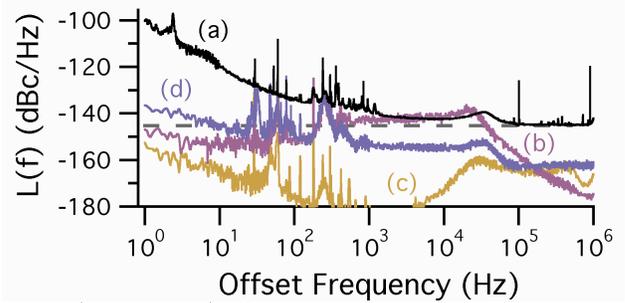

Fig. 1. (Color online) Phase noise limitation imposed by the Er:fiber-based OFD. (a) microwave phase noise on 10 GHz carrier from Fig. 2. (b) Phase noise from $f_0$ lock. (c) Phase noise from $f_b$ lock. (d) RIN converted to phase noise at the PD. The dotted horizontal line is the shot noise limit.